\RequirePackage{fix-cm}
\documentclass[smallextended]{svjour3}       % onecolumn (second format)
\smartqed  % flush right qed marks, e.g. at end of proof
\usepackage[dvips]{graphicx}
\usepackage{color}
\usepackage{graphicx}
\usepackage{epsfig}
\usepackage{amsfonts}
\usepackage{amssymb}
\usepackage{amsmath}
\usepackage{subfig}
\usepackage{bm}
\usepackage{hyperref}
\usepackage{bbm}
\usepackage{cancel}
\usepackage{color}
\usepackage[numbers,sort]{natbib}
\renewcommand\floatsep{6pt}
\renewcommand\intextsep{8pt}
\usepackage[T1]{fontenc} % if needed

\newcommand{\Res}{ \text{Res}}
\newcommand{\updown}[2]{{ \makebox [0 pt ]{$ {\scriptscriptstyle  { #1 \atop #2 } } $}  }}

\definecolor{light-gray}{gray}{0.3}
\definecolor{deep-gray}{gray}{0.5}

\newcommand{\newref}[1]{(\ref{#1})}

\begin{document}
\title{BCJ Numerators from Differential Operator of Multidimensional Residue}
\author{Gang Chen,  Tianheng Wang}


\institute{
G. Chen \at Department of Physics, Zhejiang Normal University, China\\
Department of Physics and Astronomy, Uppsala University, Sweden \\
Centre for Research in String Theory, School of Physics and Astronomy, Queen Mary University of London, U.K.
%  \\
%  \emph{Present address:} of F. Author  %  if needed
\and
T. Wang
\at
Department of Physics, Nanjing University, China  \\
Department of Physics and Astronomy, Uppsala University, Sweden \\
Institut f\"ur Physik, Humboldt-Universit\"at zu Berlin, Germany
}

\date{Received: date / Accepted: date}
\maketitle

\begin{abstract}
In previous works, we devised a differential operator for evaluating typical integrals appearing in the Cachazo-He-Yuan (CHY) forms and in this paper we further streamline this method. We observe that at tree level, the number of free parameters controlling the differential operator depends solely on the number of external lines, after solving the constraints arising from the scattering equations. This allows us to construct a reduction matrix that relates the parameters of a higher-order differential operator to those of a lower-order one. The reduction matrix is theory-independent and can be obtained by solving a set of explicitly given linear conditions. The repeated application of such reduction matrices eventually transforms a given tree-level CHY-like integral to a prepared form. We also provide analytic expressions for the parameters associated with any such prepared form at tree level. We finally give a compact expression for the multidimensional residue for any CHY-like integral in terms of the reduction matrices. We adopt a dual basis projector which leads to the CHY-like representation for the non-local Bern-Carrasco-Johansson (BCJ) numerators at tree level in Yang-Mills theory. These BCJ numerators are efficiently computed by the improved method involving the reduction matrix.
%In this paper, we propose a systematic way to evaluate the Cachazo-He-Yuan (CHY) integral at tree level. Adopting the differential operator designed for evaluating the CHY forms in our previous work, we observe that the number of independent coefficients in a given differential operator depends solely on the number of external lines. Based on this observation, the evaluating the CHY integral is translated to the decomposition of the corresponding differential operator into the widely-used minimal basis.  The decomposition is rendered highly efficient by employing the reduction matrices multiple times. We get a factorized form for the value of the CHY integral. This method is theory-independent and for Yang-Mills theory in particular, the we use the dual basis to get a CHY-like form for the non-local BCJ numerators in the minimal basis. After  evaluating the CHY-like integral in our method, the BCJ numerators are of  the manifest gauge invariance in $(n-2)$ legs for an $n$-point amplitude.
\end{abstract}


\date{\today}
\maketitle
%\tableofcontents

\section{Introduction}\label{sec:intro}
Scattering amplitudes in a number of theories can be packaged in the compact expressions known as the Cachazo-He-Yuan (CHY) forms \cite{Cachazo:2013hca,Cachazo:2013iea,Cachazo:2014nsa}. The CHY forms are originally proposed for tree-level scattering amplitudes and later generalized to loop levels~\cite{Casali:2015vta,Geyer:2015bja,Geyer:2015jch,Geyer:2016wjx,Cachazo:2015aol,He:2015yua,Feng:2016nrf,Gomez:2017lhy,Gomez:2017cpe,He:2017spx}. In the CHY form, the scattering amplitude is represented as a contour integral around the solutions to the scattering equations \cite{Cachazo:2013hca,Cachazo:2013iea,Cachazo:2014nsa}, which can be transformed to a polynomial form \cite{Dolan:2013isa,Dolan:2014ega}.  Such contour integrals can be evaluated using the integration rules and the cross-ratio method at tree and loop levels \cite{Baadsgaard:2015voa,Baadsgaard:2015hia,Cardona:2016gon,Huang:2017ydz}. Systematic methods for computing these integrals are based on the analysis of multidimensional residues on the isolated solutions of the scattering equations. One method for computing multidimensional residues involving the Groebner basis or the H-basis is discussed in \cite{Sogaard:2015dba,Bosma:2016ttj}. A useful \textsf{Mathematica} package for computing such residues is given in \cite{Larsen:2017aqb}.  

In \cite{Chen:2016fgi,Chen:2017edo} Cheung, Xu and the current authors proposed a method for evaluating the CHY forms using a differential operator and studied the combinatoric properties of the scattering equations. This method bypasses the need for solving the scattering equations and leads to the analytic evaluation of a  particular class of CHY forms, called the prepared forms.  In this paper, we further streamline the method at tree level by relating a generic CHY-like expression to a prepared form. A crucial observation in our approach is that the number of independent parameters appearing in such a differential operator is always $(n-3)!$ where $n$ is the number of external lines, regardless of the order of the operator. As will become clear in later discussions, this observation allows us to develop a method that relates higher-order differential operators with lower-order ones, through the reduction matrices for each factor of the terms in the Pfaffian expansion. Our method is theory independent and it maintains the factorized form of CHY integrand. Due to the two advantages, the CHY integral is evaluated efficiently.

As a particular application of our method, we study the construction of the Bern-Carrasco-Johansson (BCJ) numerators \cite{Bern:2008qj} from the CHY forms. The color-kinematic duality is found to hold in a number of theories \cite{Bern:2008qj,Monteiro:2011pc,Broedel:2012rc,Bargheer:2012gv,Ochirov:2013xba,Chiodaroli:2013upa,Stieberger:2014cea,Johansson:2014zca,Chiodaroli:2014xia,Huang:2013kca,Chen:2013fya,Chen:2014dfa} and extensive studies have been dedicated to computing the BCJ numerators \cite{BjerrumBohr:2010hn,Mafra:2011kj,Fu:2012uy,Mafra:2015vca, Bjerrum-Bohr:2016axv}. In \cite{Mafra:2011kj,Mafra:2016ltu}, the twistor string theory have been studied to extract the local Bern-Carrasco-Johansson (BCJ)  \cite{Bern:2008qj} numerators. The CHY forms can also be used to study the BCJ numerators and in \cite{Du:2017kpo} the local BCJ numerators are constructed. In this paper, we extract the non-local BCJ numerators in the minimal basis at tree level from the CHY forms, by introducing a dual basis projector. This way the BCJ numerators also take CHY-like expressions and can be easily studied using the differential operator and the reduction matrix.   %a closed form  for the non-local BCJ numerator by introducing the dual basis for the color-part Park-Taylor factor. This is equivalent to the closed form of BCJ numerator by the product of KLT matrix and the color-ordered amplitude. We also evaluate the CHY-like integral systematically and obtain a compact form of the non-local BCJ numerator. In this compact form, the gauge invariance for the $n-2$ legs  is obvious. 

\section{Preliminary: review of differential operator method}

Here we briefly summarize our method for computing the multidimensional residue. Let $g_1$, $g_2$, ..., $g_k$ be homogeneous polynomials in complex variables $z_1$, $z_2$, ... $z_k$. If their common zeros lie on a single \textit{isolated} point $p$ { (for homogeneous polynomials, the point $p$ is the origin)}, for a holomorphic function $\mathcal R (z_i)$ in a neighborhood of $p$, we conjecture that a differential operator $\mathbb D$ computes the residue of $\mathcal R$ at $p$ as follows
\begin{align}
\Res_{\{(g_1),\cdots,(g_k)\},p}[\mathcal{R}]
\equiv\oint \frac{dz_1\wedge \cdots \wedge dz_k}{g_1\cdots g_k} \mathcal R 
=\left. \mathbb{D}^{(m)}[\mathcal R]\right|_{z_i\rightarrow 0}, 
\end{align}
where $\mathbb D^{(m)}$ is a differential operator of order-$m$ and takes the following form,
\begin{align}
\mathbb{D}^{(m)} 
=\sum_{\{r_i\}_m}a_{r_1, r_2 ,\cdots, r_{k}}
 \partial^{r_{1}, r_2, \cdots, r_k}\,.\label{defD}
\end{align}
Here $\partial^{r_1, r_2, \cdots, r_k}=(\frac{\partial}{\partial z_1})^{r_1}(\frac{\partial}{\partial z_2})^{r_2}\cdots (\frac{\partial}{\partial z_k})^{r_k}$ and $r_i$'s are non-negative integers satisfying the Frobenius equation
$\sum_{i=1}^{k} r_i= m \equiv\sum_{i=1}^k \text{deg}(g_i) -k$. The coefficients $a_{r_1,r_2,\dots,r_k}$ are constants independent of $z_i$'s, determined uniquely by two sets of constraints arising from: 1. the local duality theorem \cite{hartshorne2013algebraic} and 2. the intersection number of the divisors $D_i=(g_i)$. Detailed discussions on these constraints can be found in \cite{Chen:2016fgi}. This conjecture applies to any such multidimensional residues around an isolated pole.

The CHY form for a tree-level scattering amplitude or a loop-level integrand is an integral on a Riemann sphere completely localized by the scattering equations. Equivalently, it is a multi-dimensional residue around the common zeros of the scattering equations. 
%Schematically, the tree-level CHY form is given by
%\begin{align}\label{eq:CHY}
%\mathcal A_n = {\color{red}\oint_{f_2=\cdots =f_{n-2}=0} {d\sigma_2 \wedge \cdots \wedge d\sigma_{n-2} \over f_2\cdots f_{n-2}} \mathcal I_n (\sigma)} \,,%~ \text{PT} (12\cdots n) ~\text{Pf}\,' (\Psi^{ij}_{ij} ) \,, 
%\end{align}
%where $f_i$'s denote the scattering equations and the integrand $\mathcal I_n(\sigma)$ is specific to the underlying theory. Here we have already taken care of the $SL(2,\mathbb C)$ conformal symmetry by fixing $\sigma_1\rightarrow0, \sigma_{n-1} \rightarrow 1,\sigma_{n}\rightarrow \infty$. 
%where %$\mathcal A_n(1,2,\cdot,n)$ denotes the $n$-point color-ordered amplitude, $
%$\text{PT}(12\cdots n)$ denotes the Parke-Taylor factor for this color ordering, 
%\begin{align}
%\text{PT}(12\cdots n) = {1 \over (\sigma_1-\sigma_2) \cdots (\sigma_{n-1}-\sigma_n) (\sigma_n-\sigma_1)}\,.
%\end{align}
%The reduced Pfaffian $\text{Pf}\,'(\Psi)$ is the \textit{only} theory-specific part in the CHY form given in the original papers \cite{Cachazo:2013hca,Cachazo:2013iea,Cachazo:2014nsa}. 
%, but for brevity we do not copy them here, since these expressions will not be used in this paper. 
The tree-level scattering equations for $n$ external particles read
\begin{align}\label{eq:originalSE}
\sum_{j\neq i} {k_i\cdot k_j \over \sigma_i -\sigma_j} = 0\,,~~~i\in[2,n-2]\,,
\end{align}
where we have already taken care of the $SL(2,\mathbb C)$ conformal symmetry by fixing $\sigma_1\rightarrow0, \sigma_{n-1} \rightarrow 1,\sigma_{n}\rightarrow \infty$. 
A simple transformation found in~\cite{Dolan:2013isa,Dolan:2014ega} takes (\ref{eq:originalSE}) to the polynomial ones
\begin{align}\label{eq:polySE}
h_t = \left.\left(\sum_{  2\leqslant i_1<i_2 <\cdots < i_t \leqslant n-1  } s_{i_1i_2\cdots i_t n}\sigma_{i_1}\cdots\sigma_{i_t} \right)\right|_{\sigma_{n-1}\rightarrow \sigma_0}\,,  ~~~t\in [1,n-3]\,,
\end{align}
where $s_{i_1\cdots i_t n} = {1\over 2}(k_{i_1}+\cdots + k_{i_t} +k_n)^2$. Here we have introduced an auxiliary variable $\sigma_0$, which fomally renders the polynomials homogeneous for the above differential operator to apply. At $\sigma_0\rightarrow 1$, the two versions of scattering equations, (\ref{eq:originalSE}) and (\ref{eq:polySE}), are equivalent.\footnote{The polynomial scattering equations are equivalent to the original ones at tree level. At loop levels, there are extra solutions, which is beyond the scope of this paper.} The Jacobian of the above transformation is given by a Vandermonde determinant $ J_n(\sigma)=\prod\limits_{1\leq r< t\leq n-1} (\sigma_t-\sigma_r)$. 

Adopting the polynomial scattering equations, a tree-level $n$-point amplitude is schematically given by a combination of the following CHY-like integrals
\begin{align}\label{eq:CHYgeneral}
\mathcal I_n(P,h_0) &=\oint\limits_{h_1=\cdots =h_{n-3}=\sigma_0-1=0} {d\sigma_2 \wedge \cdots \wedge d\sigma_{n-2} \wedge d\sigma_0 \over h_1\cdots h_{n-3} (\sigma_0-1)} {P (\sigma) \over h_0(\sigma)} \,,
 %&= {\color{red}-\sum_i \oint_{h_1=\cdots =h_{n-3}=\mathcal D_i=0} {d\sigma_0d\sigma_2 \wedge \cdots \wedge d\sigma_{n-2} \over h_1\cdots h_{n-3} \mathcal D_i(\sigma)} {\mathcal N_i (\sigma) \over (\sigma_0-1)}}
\end{align}
where the integrand is a rational function specific to the underlying theory. Its explicit expressions in different contexts can be found in \cite{Cachazo:2013hca,Cachazo:2013iea,Cachazo:2014nsa}. For our purpose, we only note that $h_0(\sigma)$ is a homogeneous polynomial and factorizes into products of degree-one polynomials. In addition to the $(n-3)$ polynomial scattering equations, $\sigma_0-1=0$ is also imposed to localize the auxiliary variable.

%We can use the global residue theorem such that the residue is taken around the solution of $h_1=\cdots =h_{n-3}=h_0=0$ instead
The global residue theorem allows us to consider the residue around the solution of $h_1=\cdots =h_{n-3}=h_0=0$ instead \footnote{Poles at infinity can in principle exist. Detailed discussions on poles at infinity are given in \cite{Chen:2016fgi}.}
\begin{align}\label{eq:CHYgeneral2}
\mathcal I_n(P,h_0) &=-\oint\limits_{h_1=\cdots =h_{n-3}=h_0=0} {d\sigma_2 \wedge \cdots \wedge d\sigma_{n-2} \wedge d\sigma_0 \over h_1\cdots h_{n-3} h_0(\sigma)} {P (\sigma) \over (\sigma_0-1)} \,.
 %&= {\color{red}-\sum_i \oint_{h_1=\cdots =h_{n-3}=\mathcal D_i=0} {d\sigma_0d\sigma_2 \wedge \cdots \wedge d\sigma_{n-2} \over h_1\cdots h_{n-3} \mathcal D_i(\sigma)} {\mathcal N_i (\sigma) \over (\sigma_0-1)}}
\end{align} The aforementioned differential operator then applies to (\ref{eq:CHYgeneral2}) as follows
\begin{align}\label{eq:ResSE}
\mathcal I_n(P,h_0)  = -\left.\left[ \mathbb D^{(m)}_{h_0} {P (\sigma) \over (\sigma_0-1)} \right] \right|_{\sigma \rightarrow 0}\,,
\end{align}
where $\sigma \rightarrow 0$ is simply a shorthand for $\sigma_r\rightarrow0, r\in\{2,\cdots, n-2, 0\}$. Namely all $\sigma_i$'s are taken to zero after the action of the differential operator. %denotes all the variable $\sigma_r, r\in \{2,\cdots, n-2, 0\}$  tend to zero after performing the differential operations. 
The differential operator $\mathbb D^{(m)}_{h_0}$  takes the form given in (\ref{defD}) with the parameters $a_{r_2, r_3,\cdots, r_{n-2},r_0}$ and $$\partial^{r_2, r_3, \cdots, r_{n-2}, r_0}=(\frac{\partial}{\partial \sigma_2})^{r_2}(\frac{\partial}{\partial \sigma_3})^{r_3}\cdots (\frac{\partial}{\partial \sigma_{n-2}})^{r_{n-2}} (\frac{\partial}{\partial \sigma_{0}})^{r_{0}}$$.
The parameters $a_{r_2, r_3,\cdots, r_{n-2}, r_0}$ are determined by the polynomial scattering equations $h_j~(j\in [1,n-3])$ and $h_0$. 
As the scattering equations are universal for all CHY-like integrals, we only specify $h_0$ in the labels of the differential operator. The order of this operator is again $m = 0+1+\cdots + (n-4) + d_{h_0}-1$, with $d_{h_0}$ denoting the degree of $h_0$.
%We omit the scattering equations and only keep $h_0$ in the down index of $\mathbb D$ because the scattering equations are universal in this paper.  The up index denote the degree of the differential operator $m = 0+1+\cdots + (n-4) + d_{h_0}$ with $d_{h_0}$ denoting the degree of $h_0$. 
We note that the scattering equations $h_j=0~(j\in [1,n-3])$ and $\sigma_0-1=0$ have multiple common solutions. Namely $(\ref{eq:CHYgeneral})$ has multiple poles. The differential operator in (\ref{eq:ResSE}) gives the sum of (\ref{eq:CHYgeneral}) evaluated at each solution, by the global residue theorem. For details, see \cite{Chen:2016fgi}
%We find that the BCJ numerators can be collected via a decomposition of the CHY form into a basis of integrals, which conceptually is similar to the aforementioned decomposition that string world sheet integrals undergo. 
% Recalling that the CHY form mathematically amounts to a multidimensional residue around the common zeros of the polynomial scattering equations,
%the differential operator for evaluating such multidimensional residues proposed in our previous papers \cite{Chen:2016fgi,Chen:2017edo} comes in handy for evaluating and manipulating the CHY forms. 
In particular, when $h_0=\sigma_i$, the CHY-like integral is studied in \cite{Chen:2017edo} and the corresponding differential operator is worked out analytically.

\section{\label{sec:RedMat} Reduction matrix and evaluation of CHY integrals}
In this section, we propose a method for relating two differential operators associated with two CHY-like integrals. To be more precise, the two CHY-like integrals of the form (\ref{eq:CHYgeneral}) share the numerator $P(\sigma)$ and their respective $h_0$ and $h'_0$ are related as $h_0(\sigma) = h'_0(\sigma) q(\sigma)$ with $q(\sigma)$ being also a polynomial. In this case, the $a$-coefficients in the two corresponding differential operators can be related by a matrix, which we call the \textit{reduction matrix}. 
This leads to a systematic evaluation of any tree-level CHY-like integral, which is given in a factorized form.

\subsection{Canonical coefficients in differential operator}\label{sec:Canon}
Consider the differential forms below
\begin{align}
\Gamma &= \frac{P(\sigma) d\sigma_2 \wedge\cdots \wedge d\sigma_{n-2} \wedge d\sigma_0}{h_1 \cdots h_{n-3} h_0}\,, 
%\Gamma' & = \frac{P(\sigma) d\sigma_2 \wedge\cdots \wedge d\sigma_{n-2} \wedge d\sigma_0}{h_1 \cdots h_{n-3} h'_0}\,,
\end{align}
 whose residue at the origin is the same as the CHY-like integral of the form (\ref{eq:CHYgeneral}).
%where $h_0 (\sigma) = h'_0 (\sigma) q(\sigma)$. Let us denote their associated differential operators as $\mathbb D_{h_0}$ and $\mathbb D_{h'$
Recall that the corresponding differential operator $\mathbb D^{(m)}_{h_0}$ takes the form of (\ref{defD}) as follows
\begin{align}\label{eq:defD}
\mathbb D^{(m)}_{h_0} = \sum_{ \{r\}_{m} } a_{r_2,\cdots, r_{n-2}, r_0} \partial^{r_2, \cdots, r_{n-2}, r_0}\,.
\end{align}
 %$\partial^{s_2 \cdots s_{n-2} s_0} = \partial_2^{s_2} \cdots \partial_{n-2}^{s_{n-2}} \partial_0^{s_0 }$ and $\{s\}_{\text{ord}(\mathbb D)}$ denotes the solutions to the Frobenius equation $s_2+\cdots s_{n-2}+s_0=\text{ord}(\mathbb D)$. 
Since the polynomial scattering equations $\{ h_1,\cdots, h_{n-3} \}$ are universal, we can always solve the local duality conditions \cite{Chen:2016fgi} arising from these polynomials first. %All the theory-specific information enters the computation through the factor $h_0$. 
These conditions read
\begin{align}\label{eq:LocalSE}
\mathbb D^{(m)}_{h_0} \left[ q_j(\sigma) h_j(\sigma) \right]=0\,,\quad\quad j =1,2,\cdots n-3\,,
\end{align}
where $q_j(\sigma)$ scans over all the monomials in $\sigma$'s of the degree $\text{deg}(q_j) =m - j$.  %These conditions lead the relations for the $a$-coefficients. These relations are of following closed form 
Substituting \eqref{eq:defD} into \eqref{eq:LocalSE}, we have 
\begin{align}\label{eq:LocalSEA}
	&\sum_{i_1<i_2<\cdots<i_t}\left[\prod_{l=2}^{n-1}(r_l+v_l)!\right]~ s_{i_1i_2\cdots i_t i_n} a_{r_2+v_2,r_3+v_3,\cdots, r_{n-2}+v_{n-2}, r_0+v_{0}}=0, 
\end{align}
where we always identify $r_{n-1}=r_0$ and $v_{n-1}=v_0$. The summation is taken over the subsets $\{i_1,\cdots, i_t\} \subset \{2,\cdots, n-1\}$ with $t\in [1, n-3]$. Here $r_s$'s are non-negative integers and $v_l$'s are defined as 
\begin{align}
	v_l=\left\{
	\begin{array}{c}
	1, ~~~\text{if}~ l\in \{i_1, i_2, \cdots, i_t\}\\
	0, ~~~\text{if}~ l\notin \{i_1, i_2, \cdots, i_t\}
\end{array}\right. .
\end{align}
For a given $n$, the number of the $a$-coefficients and the number of local duality conditions both grow as $d_{h_0}$ increases. 
However, we observe that the number of independent $a$-coefficients after solving the equations \eqref{eq:LocalSEA} is always $(n-3)!$, regardless of $m$.\footnote{This is an observation from a large number of examples, both analytic and numeric. We don't have a proof for this observation at the moment.} This allows us to choose $(n-3)!$ $a$-coefficients as a basis and expand the rest on this basis.

For the purpose of this paper, we find a particularly convenient basis choice as follows, %, and we would like to call these $a$-coefficients chosen as the parameters the \textit{canonical coefficients}. 
%Let $d_{h_0}$ denote the degree of the polynomial $h_0$ %is  and the order of the relevant differential operator is
%\begin{align}
%\text{ord}(\mathbb D) = 1+2+\cdots + (n-3) + d_{h_0} - (n-2) \,.%= 1+2+\cdots+ (n-4) + \left( d_{h_0}-1\right) \,.
%\end{align}
%We define the canonical coefficients as the ones of the following form
\begin{align}\label{eq:aCanon}
\{ a_{\gamma(0),\cdots, \gamma(n-4), (d_{h_0}-1)} |\gamma\in S_{n-3} \} \,,
\end{align}
where $S_{n-3}$ denotes the permutations. Throughout this paper, the $a$-coefficients in the above set is called the \textit{canonical coefficients}. 
The differential operator can then be rewritten only in the canonical coefficients
\begin{align}\label{eq:decomposeD}
\mathbb D^{(m)}_{h_0} = \sum_{i=1}^{(n-3)!} a_{\gamma_i(0), \cdots, \gamma_i(n-4), d_{h_0}-1} \mathcal D_i^{(m)}\,,
\end{align}
where each $\gamma_i$ denotes a different permutation. The non-canonical $a$-coefficients are expanded into the canonical ones
\begin{align}\label{eq:abasisExpansion}
& a_{r_2,\cdots, r_{n-3}, r_0} = \sum_{i=1}^{(n-3)!} c^{r_2, \cdots, r_{n-3}, r_0}_i \, a_{\gamma_i(0), \cdots,  \gamma_i(n-4), (d_{h_0}-1)}\,,
\end{align}
with the coefficients $c^{r_2, \cdots, r_{n-3}, r_0}_i$ obtained by solving \newref{eq:LocalSEA}.  Collecting the canonical coefficients, we have
\begin{align}\label{eq:calD}
\mathcal D_i^{(m)} = \partial^{\gamma_i(0),\cdots, \gamma_i(n-4) ,(d_{h_0}-1)} + \sum_{\substack{\sum r_j =m, \\ \{r_j\} \notin  S_{n-3}}} c^{r_2, \cdots, r_{n-3}, r_0}_i\, \partial^{r_2,\cdots, r_{n-2}, r_0}\,.
\end{align}
%The coefficients $c^{s_2 \cdots s_{n-3} s_0}_\gamma$ in the summation are obtained by solving the local duality constraints related to the scattering equations \newref{eq:LocalSE} and can be functions of the external momenta only. 
%Note that these coefficients are universal for all the differential operators of the same order, for a given $n$. Hence we have
%\begin{align}\label{eq:decomposeD}
%\mathbb D^{(m_0)} =& \,\left(\mathcal D_{\gamma_1}^{(m_0)},\cdots, \mathcal D_{\gamma_{M}^{(m_0)}} \right)\cdot \left[ a_{\gamma_1(0)\cdots \gamma_1(n-4) (d_0-1)} \mathbf e_1 + \cdots a_{\gamma_M(0)\cdots \gamma_M(n-4) (d_0-1)} \mathbf e_M\right] \nonumber \\ 
%=& \; a_{\gamma_1(0)\cdots\gamma_1(n-4)(d_0-1)} \mathbb D_{\mathbf b_1}+\cdots +a_{\gamma_M(0)\cdots\gamma_M(n-4)(d_0-1)} \mathbb D_{\mathbf b_M}\,,
%\end{align}
%The operators $\mathcal$
In \newref{eq:decomposeD}, the differential operator $\mathbb D^{(m)}_{h_0}$ is expanded into the basis spanned by $\mathcal D_{i}^{(m)} , i\in [1, (n-3)!]$. We note that $\mathcal D_i^{(m)}$'s are determined solely by the order $m$ and the scattering equations, independent of the actual form of $h_0$. %which we called canonical basis. From the polynomial scattering equation is of combination property, the basis is obtained in a reductive way effectively as shown in \cite{Chen:2017edo}. The closed form of these differential operators will leave to future work. %We call it ``the unit basis'' as {\color{red} the coefficient of $\mathcal D_i^{(m_0)}$) }%there is one canonical coefficient in this operator (namely the coefficient of $\partial^{\gamma_i(0)\cdots \gamma_i(n-4) (d_0-1)}$) whereas all the other canonical coefficients are zero. Therefore the canonical $a$-coefficients in $\mathcal D^{(m_0)}_i$ form a unit vector with the $i$-th component being 1.

 \subsection{Reduction matrix}

We now study the relation between the higher- and lower-order operators. Consider the meromorphic forms $\Gamma$ and $\Gamma'$ below%, the latter taking the form of the former only with $h_0$ substituted by $h'_0$, and their corresponding differential operators are $\mathbb D^{(m)}_{h_0}$ and $\mathbb D^{(m-d_q)}_{h_0'}$. 
\begin{align}\label{eq:GandGp}
\Gamma = \frac{P(\sigma) d\sigma_2 \wedge\cdots \wedge d\sigma_{n-2} \wedge d\sigma_0}{h_1 \cdots h_{n-3} h_0}\,, \quad
\Gamma' = \frac{P(\sigma) d\sigma_2 \wedge\cdots \wedge d\sigma_{n-2} \wedge d\sigma_0}{h_1 \cdots h_{n-3} h'_0}\,,
\end{align}
where $h_0(\sigma) = h'_0(\sigma) q(\sigma)$ with $q(\sigma)$ also being a polynomial of degree $d_q$. Let $\mathbb D^{(m)}_{h_0}$ and $\mathbb D^{(m-d_q)}_{h'_0}$ denote their corresponding differential operators respectively. 
For an arbitrary homogeneous polynomial $P(\sigma)$ of degree $\text{deg}(P)\leqslant m-d_q$, we must have
\begin{align}\label{eq:Drelation}
\left.\left[\mathbb D^{(m)}_{h_0} { q(\sigma) P(\sigma) \over \sigma_0-1} \right] \right|_{\sigma \rightarrow0}= \left.\left[\mathbb D^{(m-d_q)}_{h'_0}  {P(\sigma)\over \sigma_0-1} \right]\right|_{\sigma \rightarrow0}\,.
\end{align}
Plugging in the solutions of the respective non-canonical $a$-coefficients on both sides and expressing both differential operators in terms of their canonical coefficients only, the equation above becomes
\begin{align}\label{eq:arelation}
&  \sum_{i=1}^{(n-3)!} a_{\gamma_i(0), \cdots, \gamma_i(n-4), d_{h_0}-1} \left. \left[\mathcal D_i^{(m)}{ q(\sigma) P(\sigma) \over \sigma_0-1} \right] \right|_{\sigma \rightarrow0} \nonumber\\
=&  \sum_{i=1}^{(n-3)!} a_{\gamma_i(0), \cdots,\gamma_i(n-4), d_{h_0}-d_q-1} \left. \left[\mathcal D_i^{(m-d_q)}{ P(\sigma) \over \sigma_0-1} \right] \right|_{\sigma \rightarrow0} \,.
\end{align}
%where $d= \text{deg}(h_0)$ and the operators $\mathcal D^{(m)}_i$ is defined in \newref{eq:calD}. 
For this equation to hold for an arbitrary $P(\sigma)$, the coefficients of the surviving derivatives $\partial^{r_2,\cdots, r_{n-3}, r_0} P$ with $r_2+\cdots +r_{n-3}+r_0 < \text{deg}(P)$ must be the same on both sides. This leads to linear relations between the two sets of canonical $a$-coefficients, which can be written in the matrix form
\begin{align}\label{eq:Mdef}
\left(\begin{array}{c}
\vdots \\
a_{\gamma(0), \cdots, \gamma(n-4),  d_{h_0}-1}\\
\vdots
\end{array}\right) = M_{q(\sigma)}^{(n,m)}
\left(\begin{array}{c}
\vdots \\
a_{\gamma(0), \cdots, \gamma(n-4),  d_{h_0}-d_q-1}\\
\vdots
\end{array}\right)\,.
\end{align}
We name the matrix $M_{q(\sigma)}^{(n,m)}$ the \textit{reduction matrix}. % {\color{red}whose components are just functions of $s_{ij}$}. 
The reduction matrix depends only on the factor $q(\sigma)$ and the orders of the differential operators while it knows nothing about the specific expression of the factor $h'_0$. We note that although the reduction matrix depends on $q(\sigma)$, its entries are only functions of momenta.%Moreover, its dependence on the operator orders is quite trivial, as we will soon observe in examples.

%This reduction operation can be performed order by order, as long as the polynomial $h_0$ can be factorized into a product of degree-one polynomials. 
The reduction process can be performed repeatedly. Typically $h_0$ in a CHY form is completely factorized as $h_0 = q^{(1)} q^{(2)}\cdots q^{(d_{h_0})}$, where each $q^{(r)}=\sigma_i-\sigma_j$ is a degree-one polynomial in $\sigma$'s.
As a result, the canonical coefficients in $\mathbb D^{(m)}_{h_0}$ can eventually be related to those in an operator of order  $m _0\equiv m-(d_{h_0}-1)$. %Furthermore, the reduction matrix depends on the order $m$ in a rather simply way $M^{(n,m)}={1\over m-m_0}M^{(n,m_0+1)}$. 
For such a degree-one $q=\sigma_i-\sigma_j$, it is easy to check that (\ref{eq:arelation}) yields a simple relation between reduction matrices $M^{(n,m)}_q={1\over m-m_0}M^{(n,m_0+1)}_q$. 
For notational brevity, we define $M^{(n,m_0+1)}_q\equiv M^{(n)}_q$. Hence we have 
\begin{align}\label{eq:mto1}
\left(\begin{array}{c}
\vdots \\
a_{\gamma(0),\cdots, \gamma(n-4), d_{h_0}-1}\\
\vdots
\end{array}\right) = {M_{q^{(1)}}^{(n)} M_{q^{(2)}}^{(n)}\cdots {M_{q^{(d_{h_0}-1)}}^{(n)} }\over (d_{h_0}-1)!}\left(\begin{array}{c}
\vdots \\
a_{\gamma(0),\cdots, \gamma(n-4), 0}\\
\vdots
\end{array}\right)\,.
\end{align}
We note that the ordering of these $q^{(r)}$ factors do not affect the eventual evaluation of the CHY-like integral. The choice of $(d_{h_0} -1)$ factors for the reduction process is also irrelevant, although certain choices might be more convenient in particular cases.
%As the inverse of the reduction matrix is linear to the variable and $q^{(r)}=\sigma_{r_i}-\sigma_{r_j}$ in the CHY-like integrand, the reduction matrix is of following relations 

The inverse of the reduction matrix is linear in the variables in the subscript, namely\footnote{The relation (\ref{eq:arelation}) can be schematically rewritten as $L_q \cdot \vec{a}_{d_{h_0}-1} = R \cdot \vec{a}_{d_{h_0}-d_q-1}$ where $\vec{a}_{d_{h_0}-1}$ and $\vec{a}_{d_{h_0}-d_q-1}$ are the two column vectors on the left and right sides of (\ref{eq:Mdef}) respectively. $L_q$ and $R$ are matrices following directly from (\ref{eq:arelation}) and we note the matrix $R$ is independent of $q$. Hence $M_q = L^{-1}_q\cdot R$. For $q=\sigma_{r_1}-\sigma_{r_2}$, (\ref{eq:arelation}) yields $L_{\sigma_{r_1}-\sigma_{r_2}}=L_{\sigma_{r_1}}-L_{\sigma_{r_2}}$. Moreover, since $M_{\sigma_{r_1}-\sigma_{r_2}} = L_{\sigma_{r_1}-\sigma_{r_2}}^{-1}\cdot R$ and $M_{\sigma_{r_i}} = L_{\sigma_{r_i}}^{-1}\cdot R,~i=1,2$, taking the inverse of the three reduction matrices, we obtain the relation (\ref{eq:MinvRel}).}
\begin{align}\label{eq:MinvRel}
	M^{(n)}_{\sigma_{r_1}-\sigma_{r_2}}=\Big((M^{(n)}_{\sigma_{r_1}})^{-1}-(M^{(n)}_{\sigma_{r_2}})^{-1}\Big)^{-1}.
\end{align}
Hence we only need to construct $(M^{(n)}_{\sigma_{r}})^{-1}$, where $r\in \{2,\cdots,n-2, 0\}$. For $q(\sigma)=\sigma_0$, it is easy to verify that the reduction matrix $M^{(n)}_{\sigma_0}$ is the $(n-3)!$-dimensional identity matrix.
For $q(\sigma)=\sigma_r$, the equation (\ref{eq:arelation}) reads 
\begin{align}\label{eq:aReduction}
	 (\gamma({r-2})+1)a_{\gamma(0), \cdots, \gamma(r-3), \gamma({r-2})+1,\gamma(r-1),\cdots, \gamma(n-4), 0}
	 =a_{\gamma(0), \cdots,  \gamma(n-4), 0},
\end{align}
which holds for any $\gamma\in S_{n-3}$. The $a$-coefficient on the left-hand side above is \textit{not} a canonical coefficient and can be rewritten in terms of the canonical ones as follows
\begin{align}
& a_{\gamma(0), \cdots, \gamma(r-3), \gamma({r-2})+1,\gamma(r-1),\cdots, \gamma(n-4), 0} \nonumber\\
	=&\sum_{j=1}^{(n-3)!}\Big(c_j^{\gamma(0), \cdots, \gamma(r-3),\gamma({r-2})+1,\gamma(r-1),\cdots, \gamma(n-4), 0} a_{\gamma_j(0),\cdots, \gamma_j(n-4), 1}\Big),
\end{align}
where the coefficients $c$'s are defined in (\ref{eq:abasisExpansion}) and are solely determined by (\ref{eq:LocalSEA}). 
From this equation we read out the elements of $(M^{(n)}_{\sigma_{r}})^{-1}$ for $r\in [2,n-2]$ as follows
\begin{align}\label{eq:MSinSigma}
	(M^{(n)}_{\sigma_{r}})^{-1}_{ij}=(\gamma_i({r-2})+1)c_j^{\gamma_i(0), \cdots, \gamma_i(r-3),\gamma_i({r-2})+1,\gamma_i(r-1),\cdots, \gamma_i(n-4), 0}.
\end{align}
%Recall that the coefficients $c$'s here relate the non-canonical $a$-coefficients to the canonical ones and are only determined by the scattering equations. 
%The coefficients $c$ only relate to the scattering equations as discussed above. 
%In $h_0$, there is a factor $\sigma_i$ remaining at the end of the reduction. This factor is related to the $a$-coefficients for the prepared forms. Using the reduction matrices, any CHY-like integral with completely factorized $h_0$ can be related to such prepared forms. They are obtained as \cite{Chen:2017edo} for one-loop scattering equations.  

Using the reduction matrices repeatedly, any CHY-like integral with a completely factorized $h_0$ can be related to one with $h'_0=\sigma_r$.\footnote{Even if $h_0$ does not have a factor $\sigma_r$, one can always multiply $\sigma_r/\sigma_r$ to the integrand and apply the reduction process to other factors.} The latter is the so-called prepared form and in \cite{Chen:2017edo} such CHY-like integrals with the one-loop scattering equations are studied. 
Here we repeat the exercise for the tree-level scattering equations. Let $a^{(\sigma_r)}$ denote the $a$-coefficients in the differential operator associated with the tree-level prepared form with $h_0=\sigma_r$. We obtain the following analytical expressions for the canonical ones
\begin{align}\label{eq:valA1}
	&a_{\gamma(0),\cdots,\gamma(n-4),0}^{(\sigma_r)}=\nonumber\\
	&\left\{\begin{array}{cc}
	{\text{sgn}(\gamma)}(n-3)!\over \left.\big(\partial^{\gamma(0)+1,\cdots,\gamma(r-3)+1,0,\gamma(r-1)+1,\cdots,\gamma(n-4)+1,1} (h_1h_2\cdots h_{n-3})\big)\right |_{\sigma\rightarrow 0}, &\text{for}~  \gamma(r-2)=0 \\
   0 &\text{for others} 	
   \end{array}\right. ,\nonumber\\
	&a_{\gamma(0),\cdots, \gamma(n-4), 0}^{(\sigma_0)}=\begin{array}{c}{{\text{sgn}(\gamma)}(n-3)!\over \left.\left(\partial^{\gamma(0)+1,\cdots, \gamma(n-4)+1,0} (h_1h_2\cdots h_{n-3})\right)\right |_{\sigma\rightarrow 0}}\end{array},
\end{align}
where $\text{sgn}(\gamma)$ denotes the signature of the permutation $\gamma$.
%Finally we solve the CHY integral completely and get the residues generally in the CHY integral \eqref{eq:ResSE} 
With the reduction matrices and the $a$-coefficients above, a generic CHY integral (\ref{eq:ResSE}) can be evaluated straightforwardly
\begin{align}\label{eq:ReductionRes}
&\mathcal I_n(P, h_0) = -\left. \left[\mathbb D^{(m)}_{h_0} {P (\sigma) \over (\sigma_0-1)} \right] \right|_{\sigma \rightarrow 0}={-1\over (d_{h_0}-1)!}\times\,\\
&\sum_{i,j=1}^{(n-3)!}\Big(M_{q^{(1)}}^{(n)} \cdots  M_{q^{(d_{h_0}-1)}}^{(n)}  \Big)_{ij}  a_{\gamma_j(0),\cdots,\gamma_j(n-4),0}^{(\sigma_r)}  \left.\left[\mathcal D_i^{(m)} {P (\sigma) \over (\sigma_0-1)} \right] \right|_{\sigma \rightarrow 0}.\nonumber
\end{align}

\subsection{Examples}
% We have presented a general discussion on the construction of the so-called reduction matrix. A direct application of the reduction matrix is to evaluate the CHY form analytically for general multiplicity. We will present the details in evaluating one typical CHY integral in the Yang-Mills field theory.  
Here we consider a couple of examples in detail to demonstrate the evaluation of tree-level CHY integrals, using the reduction matrix discussed above.

At four points, we have only one scattering equation after gauge fixing $\sigma_1\rightarrow0, \sigma_3\rightarrow1,\sigma_4\rightarrow\infty$. We consider the integral below as a simple example
%Let us take the following CHY integral as a simple example
  \begin{align}
 	\mathcal I_{4}(1,\sigma_2-1)=\oint_{h_1=0} {d\sigma_2 \over h_1}{1\over  \sigma_2-1}= \oint\limits_{h_1=\sigma_0-1=0}{d\sigma_2\wedge d\sigma_0 \over h_1 (\sigma_0-1)} {\sigma_0\over  (\sigma_2-\sigma_0)\sigma_0},
 \end{align}
where in the second equal sign we have homogenized the scattering equation and the original denominator of the integrand. We have also used the trick of multiplying $\sigma_0/\sigma_0$ to the integrand for the later use of the prepared form. 
The homogenized polynomial scattering equation reads
 \begin{align}
 	h_1= s_{13}\sigma_2+s_{12} \sigma_0.
\end{align}
This integral is then given by the action of a differential operator as follows
 \begin{align}
	\mathcal I_{4}(1,\sigma_2-1) =-\left.\left[ \mathbb D^{(1)}_{(\sigma_2-\sigma_0)\sigma_0} {\sigma_0 \over (\sigma_0-1)} \right] \right|_{\sigma \rightarrow 0}, 
 \end{align}
 where $\mathbb D^{(1)}_{(\sigma_2-\sigma_0)\sigma_0}=a_{1,0}\partial^{1,0}+a_{0,1}\partial^{0,1}$ and $\partial^{r_2,r_0}=({\partial\over \partial{\sigma_2}})^{r_2}({\partial\over \partial{\sigma_0}})^{r_0}$. The non-canonical $a_{1,0}$ is related to the canonical $a_{0,1}$ via (\ref{eq:LocalSEA}) and we have 
 \begin{align}
 \mathbb D^{(1)}_{(\sigma_2-\sigma_0)\sigma_0} = a_{0,1} \mathcal D^{(1)}_1, ~~~~\text{with}~~~\mathcal D^{(1)}_1=\partial^{0,1} -\frac{ s_{12}}{s_{13}}\partial^{1,0}.
 \end{align}
 %According to Eq. \eqref{eq:arelation}, the reduction matrix for the factor $(\sigma_2-\sigma_0)$ in the denominator of the integrand is 
 The reduction matrix $M^{(4)}_{\sigma_2-\sigma_0}$ can be obtained from (\ref{eq:MinvRel}) and (\ref{eq:MSinSigma}) directly, which relates $a_{0,1}$ above to $a_{0,0}^{(\sigma_0)}$ given by (\ref{eq:valA1}). We have
 \begin{align}\label{eq:Mexample4}
 	 M_{\sigma_2-\sigma_0}^{(4)}=-\frac{s_{13}}{ (s_{12}+s_{13})},~~~~~ a_{0,0}^{(\sigma_0)}=\left.\left[{1\over \partial^{1,0} (h_1)}\right]\right |_{\sigma\rightarrow 0}={1\over s_{13}}.
 \end{align}
% Finally, only a factor $\sigma_0$ is in the denominator which meet the requirement of the prepared form. According to Eq. \eqref{eq:valA1} we have 
% \begin{align}
%	a_{0,0}(\sigma_0)=\left.\left[{1\over \partial^{1,0} (h_1)}\right]\right |_{\sigma\rightarrow 0}={1\over s_{13}}.
%\end{align}
%According to \eqref{eq:ReductionRes}, we get
Hence the integral reads simply
\begin{align}
	\mathcal I_{4}(1,\sigma_2-1)=- M_{\sigma_2-\sigma_0}^{(4)}a_{0,0}^{(\sigma_0)}\left.\left[\mathcal D_1^{(1)} {\sigma_0 \over (\sigma_0-1)} \right] \right|_{\sigma \rightarrow 0}=-\frac{1}{s_{12}+s_{13}}.
\end{align}
%Similarly, the other integral is evaluated as
%\begin{align}
%	\mathcal I_{4,2}&=-\mathcal M_{(\sigma_2-\sigma_0)}^{4, 2}\mathcal M_{(\sigma_2-\sigma_0)}^{4, 1}a_{0,0}(\sigma_0)\left.\left[(\partial^{0,2} -\frac{ s_{12}}{s_{13}}\partial^{1,1} ) {1 \over (\sigma_0-1)} \right] \right|_{\sigma \rightarrow 0}\nonumber\\
%	&=\frac{s_{13}}{(s_{12}+s_{13})^2}.
%\end{align}
%This integral has the double pole which are usually an trouble in the integral rule method. Even more general integral is also calculated directly 
% \begin{align}
 %	 \mathcal I_{4,r}=\oint_{h_1=0} {d\sigma_2\over (-1 + \sigma_2)^r}=(-1)^r\frac{s_{13}^{r-1}}{(s_{12}+s_{13})^r}.
% \end{align}
  
At five points, we consider below a typical CHY-like integral with a double pole and a nontrivial numerator
\begin{align}
\mathcal I_{5}((\sigma_2-1) \sigma_3, (\sigma_3-1)^2\sigma_2) = \oint\limits_{\updown{h_1=h_2=0}{\sigma_0-1=0}}\frac{(\sigma_2-1) \sigma_3~d\sigma_2\wedge d\sigma_3\wedge d\sigma_0}{h_1 h_2 (\sigma_0-1) (\sigma_3-\sigma_0)^2\sigma_2},
\end{align}
where we have homogenized the scattering equations and the denominator $(\sigma_3-1)^2\sigma_2$ with the auxiliary variable $\sigma_0$. The numerator does not need to be homogenized for the application of our method.
The homogenized polynomial scattering equations are 
 \begin{align}
 	h_1&=s_{134}\sigma_{2} +s_{124}\sigma_3 +s_{123}\sigma_0\nonumber\\
 	h_2&=s_{14} \sigma_2 \sigma_3+s_{13}\sigma_2\sigma_0 +s_{12}\sigma_3 \sigma_0.
 \end{align}
This integral is then given by the action of the following differential operator
 \begin{align}
 	\mathcal I_{5}((\sigma_2-1) \sigma_3, (\sigma_3-1)^2\sigma_2)=-\left. \left[\mathbb D^{(3)}_{(\sigma_3-\sigma_0)^2\sigma_2} {(\sigma_2-1) \sigma_3 \over (\sigma_0-1)} \right] \right|_{\sigma \rightarrow 0}.
 \end{align}
 The differential operator takes the form below 
 \begin{align}
	\mathbb D^{(3)}_{(\sigma_3-\sigma_0)^2\sigma_2}=a_{0,0,3}\partial^{0,0,3}+a_{1,1,1}\partial^{1,1,1}+a_{0,1,2}\partial^{0,1,2}+\cdots,
\end{align}
where $\partial^{r_2,r_3,r_0}=({\partial\over \partial{\sigma_2}})^{r_2}({\partial\over \partial{\sigma_3}})^{r_3}({\partial\over \partial{\sigma_0}})^{r_0}$ and we have only written out the terms that have nonzero contributions. Here \eqref{eq:LocalSEA} gives the following equations
\begin{align}
	6s_{45}a_{0,0,3}+2s_{35}a_{0,1,2}+2s_{25}a_{1,0,2}&=0\nonumber\\
	s_{235}a_{1,1,1}+2s_{345}a_{0,1,2}+2s_{245}a_{1,0,2}&=0.
\end{align}
The canonical coefficients here are $\{a_{0,1,2} , a_{1,0,2}\}$ and hence we have
\begin{align}
	\mathbb D^{(3)}_{(\sigma_3-\sigma_0)^2\sigma_2}=a_{0,1,2}\mathcal D_1^{(3)}+a_{1,0,2}\mathcal D_2^{(3)},
\end{align}
where 
\begin{align}\label{eq:DiffOPFiveHighest}
	\mathcal D_1^{(3)}&=-\frac{ s_{35}}{3 s_{45}}\partial^{0,0,3}-\frac{2  s_{12}}{s_{14}}\partial^{1,1,1}+\partial^{0,1,2}+\cdots\nonumber\\
	\mathcal D_2^{(3)}&=-\frac{ s_{2,5}}{3 s_{45}}\partial^{0,0,3}-\frac{2 s_{13}}{s_{14}}\partial^{1,1,1}+\partial^{1,0,2}+\cdots .
\end{align}

Applying the reduction matrix for the factor $\sigma_3-\sigma_0$ \textit{twice}, the above canonical coefficients $\{a_{0,1,2} , a_{1,0,2}\}$ are related to the canonical coefficients $\{a^{(\sigma_2)}_{0,1,0}, a^{(\sigma_2)}_{1,0,0}\}$.  %According to Eq. \eqref{eq:ReductionRes}, we need to construction the reduction matrix $M^{(5)}_{\sigma_3-\sigma_0}$ and a-coefficients value of $\sigma_2$ for the differential operator $\mathbb D^{(3)}_{(\sigma_3-\sigma_0)^2\sigma_2}$. The linearity property for the inverse of the reductive matrix leads to  $$M^{(5)}_{\sigma_3-\sigma_0}=\left((M_{\sigma_3}^{(5)})^{-1}-(M_{\sigma_0}^{(5)})^{-1}\right)^{-1}.$$ 
Hence we only need to construct $M^{(5)}_{\sigma_3-\sigma_0}$. 
The reduction matrix for  $\sigma_0$ is just two dimensional identity matrix. 
Recall the linearity property (\ref{eq:MinvRel}) and that $M_{\sigma_0}^{(5)}$ is simply a two-dimensional identity matrix. The only nontrivial ingredient here is $(M_{\sigma_3}^{(5)})^{-1}$. Recall that $M_{\sigma_3}^{(5)}$ is defined to relate an order-$2$ differential operator to an order-$1$ one. For $q=\sigma_3$, the relation (\ref{eq:aReduction}) reads
%The inverse of the reduction matrix for  $\sigma_3$ is read out from according to Eq. \eqref{eq:aReduction} \eqref{eq:MSinSigma}. The reduction matrix is deduced from the relation between a-coefficient of order $1$ and $2$ 
\begin{align}\label{eq:a2to1Example}
%\left(\begin{array}{cc}
%a_{0,1,0}\\ a_{1,0,0}	
%\end{array}\right)=\left(\begin{array}{cc}
%2a_{0,2,0}\\ a_{1,1,0}	
%\end{array}\right).
a_{0,1,0} = 2 a_{0,2,0},~~~~~ a_{1,0,0}=a_{1,1,0}.
\end{align}
Moreover, the local duality conditions (\ref{eq:LocalSEA}) for the $a$-coefficients in the order-$2$ differential operator read
\begin{align}\label{eq:localSEAorder2example}
	2s_{35}a_{0,2,0}+s_{25}a_{1,1,0}+s_{45}a_{0,1,1}&=0,\nonumber\\
	s_{235}a_{1,1,0}+s_{345}a_{0,1,1}+s_{245}a_{1,0,1}&=0.
\end{align}
%Then we have the inverse of reduction matrix $(M_{\sigma_3}^{(5)})^{-1}$ which relates the $a$-coefficients for the differential operators of different orders as following
Plugging (\ref{eq:a2to1Example}) in (\ref{eq:localSEAorder2example}) we obtain
\begin{align}
\left(\begin{array}{cc}
a_{0,1,0}\\ a_{1,0,0}	
\end{array}\right)=
\left(
\begin{array}{cc}
 -\frac{s_{35}}{s_{45}} & -\frac{s_{25}}{s_{45}} \\
 \frac{s_{12} s_{35}}{s_{13} s_{45}} & \frac{s_{12} s_{25}-s_{14} s_{45}}{s_{13} s_{45}} \\
\end{array}
\right)
\left(\begin{array}{cc}
a_{0,1,1}\\ a_{1,0,1}	
\end{array}\right),
\end{align}
where the matrix is $(M_{\sigma_3}^{(5)})^{-1}$. With this matrix obtained, we can compute $M_{\sigma_3-\sigma_0}^{(5)}$ straightforwardly. 
% the reduction matrix for the factor in the denominator $\sigma_2-\sigma_3$ is 
%\begin{align}
%	M^{(5)}_{\sigma_3-\sigma_0}=\left(
%\begin{array}{cc}
 %\frac{\left(s_{13}+s_{14}\right) s_{35}}{ s_{12} s_{25}- \left(s_{13}+s_{14}\right) \left(s_{35}+s_{45}\right)} & \frac{s_{13} s_{25}}{ s_{12} s_{25}-\left(s_{13}+s_{14}\right) \left(s_{35}+s_{45}\right)} \\
 %-\frac{s_{12} s_{35}}{ s_{12} s_{25}- \left(s_{13}+s_{14}\right) \left(s_{35}+s_{45}\right)} & \frac{s_{14} \left(s_{35}+s_{45}\right)-s_{12} s_{25}}{ s_{12} s_{25}- \left(s_{13}+s_{14}\right) \left(s_{35}+s_{45}\right)} \\
%\end{array}
%\right).
%\end{align}
The canonical coefficients $\{a^{(\sigma_2)}_{0,1,0}, a^{(\sigma_2)}_{1,0,0}\}$ are again given by \eqref{eq:valA1} as follows
\begin{align}
	a_{0,1,0}^{(\sigma_2)}={2\over \Big(\partial^{0,2,1} (h_1h_2)\Big)\Big |_{\sigma\rightarrow 0}}={1\over s_{1 2} s_{124}}, && a_{1,0,0}^{(\sigma_2)}=0.
\end{align}
Hence we have 
\begin{align}
	\mathcal I_{5}&=-{1\over 2}\sum_{i,j=1}^2\Big( M^{(5)}_{\sigma_3-\sigma_0}M^{(5)}_{\sigma_3-\sigma_0}\Big)_{i,j}a_{\gamma_i(0),\gamma_i(1),0}^{(\sigma_2)}\left.\left[\mathcal D^{(3)}_j {1 \over (\sigma_0-1)} \right] \right|_{\sigma \rightarrow 0}\nonumber\\
	&=\frac{s_{234} s_{14}+s_{12}s_{24}+s_{234}s_{24} }{s_{12} s_{234} s_{34}^2},
\end{align} 
where $\gamma_i\in S_2$.

\section{\label{sec:Examples}Gauge Invariant BCJ Numerators of Yang-Mills}
In previous sections, we have presented general discussions on the evaluation of tree-level CHY-like integrals, using the differential operator and the reduction matrix. This approach only hinges on the scattering equations and the factorized form of the integrand and therefore applies to all theories whose amplitudes admit the CHY representations. In this section, we consider a particular application of the method: the construction of the BCJ numerators for tree-level Yang-Mills amplitudes.

%\subsection{Dual basis of BCJ numerators}
\subsection{BCJ numerators from CHY form}\label{sec:DualBasis}
The CHY form \cite{Cachazo:2013hca,Cachazo:2013iea,Cachazo:2014nsa} for a color-ordered amplitude in Yang-Mills at tree level reads
\begin{align}\label{eq:AinCHY}
A(1,\alpha, n-1, n) = \oint\limits_{h_1=\cdots =h_{n-3}=0}d\Omega_n   \sigma_{1\,n-1\,n}^2\text{PT}(1\,\alpha\, n-1\,n)  \text{Pf}\,'[\Psi_{1n}(\sigma)]\, ,
\end{align}
where $\alpha$ denotes a permutation of $\{2,3,\cdots,n-2\}$. Here we have used the shorthand notation $\sigma_{i_1\cdots i_t}= (\sigma_{i_t}-\sigma_{i_1})\prod_{j=1}^{t-1}(\sigma_{i_j}-\sigma_{i_{j+1}})$ and introduced the measure
\begin{align}
 d\Omega_n =  { J_n(\sigma)~ d\sigma_2 \wedge \cdots \wedge d\sigma_{n-2} \over h_1\cdots h_{n-3}}\,. 
\end{align} 
Recall that $ J_n(\sigma)=\prod\limits_{1\leq r< t\leq n-1} (\sigma_t-\sigma_r)$ denotes the Jacobian of the transformation to the polynomial scattering equations. We have adopted the gauge-fixing $\sigma_1\rightarrow 0, \sigma_{n-1}\rightarrow 1, \sigma_n \rightarrow\infty$ and $\sigma_{1\,n-1\,n}^2$ is the factor introduced after the gauge-fixing.
% and introduced the notation for the measure {\color{magenta} maybe move to earlier sections}
%\begin{align}
%	J(\sigma)=\prod_{1\leq r< t\leq n-1} (\sigma_t-\sigma_r)
%\end{align}
%\begin{align}
%J(\sigma)=Det\left(
%\begin{array}{ccccc}
% 1 & 1 & \cdots &1 & 1 \\
 %\sigma_1 & \sigma_2 &  \cdots &\sigma_{n-2} & \sigma_{n-1} \\
  %\vdots & \vdots &  \ddots &\vdots & \vdots \\
 %\sigma_1^{n-3} & \sigma_2^{n-3} &\cdots & \sigma_3^{n-3} & \sigma_4^{n-3} \\
%\sigma_1^{n-2} & \sigma_2^{n-2} & \cdots &\sigma_3^{n-2} & \sigma_4^{n-2} \\
%\end{array}
%\right)
%\end{align}
%\begin{align}
%{\color{magenta} d\Omega_n = (\text{fill in the factors})\; {d\sigma_2 \wedge \cdots \wedge d\sigma_{n-2} \over h_1\cdots h_{n-3}}\,. }
%\end{align}
The Parke-Taylor factor corresponding to a given color ordering $(1,\alpha,n-1,n)$ reads
\begin{align}
 \text{PT}(1\,\alpha\, n-1\, n) = {1 \over \sigma_{1\,\alpha(2)\cdots \alpha(n-2)\,n-1\, n} }\,.
\end{align}
$\text{Pf}\,'[\Psi_{1n}(\sigma)]$ denotes the reduced Pfaffian of the matrix $\Psi(\sigma)$ with the first and the last columns and rows removed. The explicit expression for the reduced Pfaffian is given in \cite{Cachazo:2014nsa} and there is a freedom of removing the $i$-th and $j$-th rows and columns of $\Psi$ for any $(i,j)$. Here for simplicity, we choose $(i,j)=(1,n)$.
%we have replaced the original $(\sigma_{n-2}-1)$ resulted from the gauge-fixing by $(\sigma_{n-2}-\sigma_{0})$, rendering the denominator homogeneous. 

On the other hand, the above color-ordered amplitudes is related to the BCJ numerators as follows,
\begin{align}\label{eq:A2BCJ}
A(1,\alpha, n-1, n)=\sum_{\beta\in S_{n-3}}\mathbf m(\alpha| \beta)N(1\,\beta \, n-1\, n)\,, 
\end{align}
where $\mathbf m(\alpha| \beta)$ denotes the propagator matrix \cite{Vaman2010}, whose rows and columns are labeled by the color orderings $(1,\alpha, n-1 ,n)$ and $(1,\beta, n-1 ,n)$. The BCJ numerators $N(1\,\beta \, n-1\, n)$ are in the minimal basis. 
Similar to the color-ordered amplitude, the propagator matrix also admits a CHY-like representation
\begin{align}\label{eq:minCHY}
 \mathbf m(\alpha| \beta)
= -\oint\limits_{h_1=\cdots =h_{n-3}=0 } \sigma^2_{1\,n-1\,n}\text{PT}(1 \,\alpha\, n-1\, n) \text{PT}(1\,\beta \,n-1 \, n)  d\Omega_n\,  .
 \end{align} 
Thus, comparing (\ref{eq:AinCHY}) and (\ref{eq:minCHY}), we see that if $\text{Pf}\,'[\Psi_{1n}(\sigma)]$ can be expanded in the basis spanned by the Parke-Taylor factors $\{ \text{PT}(1\,\beta \,n-1 \, n) | \beta \in S_{n-2} \} $ with $\sigma$-independent coefficients, these coefficients are the BCJ numerators. To extract the BCJ numerators, we adopt a \text{dual basis projector} $\overline{\text{PT}}(1\,\alpha\, n-1\, n)$ that satisfies the condition
 \begin{align}\label{eq:dualBasis}
\oint\limits_{h_1=\cdots =h_{n-3}=0 } (-1)\sigma^2_{1\,n-1\,n}\overline{\text{PT}}(1 \,\alpha\, n-1\, n) \text{PT}(1\,\beta \,n-1 \, n)   d\Omega_n =\delta_{\alpha\beta}.
 \end{align} 
%\begin{align}
%\langle \overline{\text{PT}}_{1\alpha n-1 n}, \text{PT}(1\beta n-1 n)\rangle_{CHY} =\delta_{\alpha\beta}\, ,
%\end{align}
Recall that the propagator matrix $\mathbf m(\alpha|\beta)$ here is a $(n-3)!\times (n-3)!$ square matrix. In this case, $\mathbf m(\alpha|\beta)$ is invertible and its inverse is the KLT matrix $\mathcal S_{\alpha\beta}$ \cite{Kawai:1985xq,BjerrumBohr:2010hn,Bern:1998sv}. Thus the following is obviously the solution of the dual basis projector,
\begin{equation}
\label{eq:ProFac}
\overline{\text{PT} }({1\,\alpha\, n-1\,n})=\sum_{\beta}\mathcal{S}_{\alpha\beta}\text{PT}(1\,\beta\, n-1\, n)\,.
\end{equation}
Using the dual basis,  the BCJ numerator is given as a CHY-like integral
\begin{align}\label{eq:NuminCHY}
N(1\,\alpha\, n-1\, n) = \oint\limits_{h_1=\cdots =h_{n-3}=0} d\Omega_n \,\overline{\text{PT}}(1\,\alpha\, n-1\, n)  \sigma_{1n-1n}^2 \text{Pf}\,'[\Psi_{1n}(\sigma)] \,,
\end{align}
The CHY-representations of BCJ numerators can be easily evaluated using the differential-operator based method. 
Generally speaking, such BCJ numerators in the minimal basis are all non-local. They may contain poles that do not correspond to the propagators in trivalent diagrams. But these unphysical poles do not contribute to the amplitudes. As we will observe in examples, the numerators computed this way are gauge invariant in $n-2$ legs and respect the crossing symmetry under the permutation of $n-3$ legs.

\subsection{Yang-Mills BCJ numerator at four and five points}
In this section, we construct the BCJ numerators at four and five points to further illustrate the application of the differential operator and the reduction matrix. As discussed above, the BCJ numerator can be computed by a CHY-like integral, which is readily evaluated by our differential-operator based method.

%As discussed above, the CHY form of the BCJ numerator is obtained from the amplitude with the color Park-Taylor factor be replaced by the dual color factor. The CHY form of the BCJ numerator is 
At four points, the minimal basis consists of only one independent BCJ numerator, which we choose to be $N(1234)$. In this case, (\ref{eq:NuminCHY}) gives
\begin{align}
\label{eq:AmpYM4}
 N(1234) = \oint_{h_1=0}\sigma_{134}^2\overline{\text{PT}}(1234) \text{Pf}\,'[\Psi_{14}(\sigma)] d\Omega_4 \,,
\end{align}
where only one integration variable $\sigma_2$ is left after gauge fixing $(\sigma_1\rightarrow 0, \sigma_3\rightarrow 1,\sigma_4\rightarrow\infty)$. %Recall that the homogenized scattering equation reads $h_1 = s_{13}\sigma_2+s_{12} \sigma_0=0$. 
The KLT matrix is given in \cite{BjerrumBohr:2010hn} and this leads to the dual basis projector below 
\begin{align}\label{eq:dualProj4}
	\overline{\text{PT}}(1234) =\frac{s_{12} s_{1,23}}{s_{13}}{\text{PT}}(1234). 
\end{align}
The reduced Pfaffian can be written as the following \cite{Lam:2016tlk}
\begin{align}\label{eq:redPfYM4}
\text{Pf}\,'[\Psi_{14}(\sigma)]=&- {\epsilon_1{\cdot} F_{2}{\cdot} F_{3}{\cdot} \epsilon_4\over \sigma_{1234}}-{\epsilon_1{\cdot} F_{3}{\cdot} F_{2}{\cdot} \epsilon_4\over \sigma_{1324}} +{\epsilon_1{\cdot} \epsilon_4\text{tr}( F_{2}{\cdot}F_{3})\over 2\sigma_{14}^2\sigma_{23}^2} \nonumber\\
&+{\epsilon_1{\cdot} F_{2}{\cdot} \epsilon_4\over \sigma_{124}}C_{(3)}+{\epsilon_1{\cdot} F_{3}{\cdot} \epsilon_4\over \sigma_{134}}C_{(2)} -{\epsilon_1{\cdot} \epsilon_4\over \sigma_{14}^2}C_{(2)}C_{(3)} \,,
\end{align}
where we have adopted the following notations also used in \cite{Lam:2016tlk}
\begin{align}\label{eq:defBrackets}
 F_{i}=k_i^{\mu}\epsilon_i^\nu-\epsilon_i^{\mu}k_i^\nu\,,&&C_{(i)}=\sum_{j=1,j\neq i}^4 {k_j\cdot\epsilon_i\over \sigma_j-\sigma_{i}}\,.
\end{align}
%Since $F_i\rightarrow0$ as $\epsilon_i\rightarrow k_i$, any term containing $F_i$ is manifestly gauge invariant in leg $i$. Therefore the first line in (\ref{eq:redPfYM4}) is manifestly gauge invariant in leg $2$ and leg $3$. When $\epsilon_i\rightarrow k_i$,  $C_{(i)}$ vanishes on the scattering equations, and thus is also gauge invariant in leg $i$. Therefore the second line in (\ref{eq:redPfYM4}) is also gauge invariant in leg $2$ and leg $3$. The gauge invariance in leg $1$ and leg $4$, although not manifest, can be verified when all the terms are summed up. 
%To get a compact form for the BCJ numerator, we also require the obvious gauge invariance for the unfixed legs even after evaluating the CHY integral. By using the reduction matrix, this requirement is easy to be realized. As an example, we consider the term ${\epsilon_1{\cdot} F_{2}{\cdot} \epsilon_4\over \sigma_{124}}C_{(3)}$ 
We now demonstrate the evaluation of (\ref{eq:AmpYM4}) using the reduction matrix. We consider the term below as an example and all other terms can be computed in the same way,
\begin{align}
\label{eq:AmpYM4EX1}
 N_4\equiv \oint\limits_{h_1=\sigma_0-1=0}\sigma_{134}^2\, \overline{\text{PT}}(1234) \,{\epsilon_1{\cdot} F_{2}{\cdot} \epsilon_4\over \sigma_{124}}\,C_{(3)}\, {d\Omega_4} \,.
\end{align}
Plugging in (\ref{eq:dualProj4}) and homogenizing the scattering equations and the denominators, we have 
 \begin{align}
 	N_4&=\frac{s_{12} s_{1,23}\epsilon_1{\cdot} F_{2}{\cdot} \epsilon_4}{s_{13}}\oint\limits_{\updown{h_1=0}{\sigma_0-1=0}}\frac{d\sigma_2\wedge d\sigma_0}{(\sigma _{2}-\sigma_0)(\sigma _{0}-1)}\left(\frac{\epsilon_3{\cdot}k_1 }{\sigma_0}-\frac{\epsilon_3{\cdot}k_1}{\sigma_{2}}-\frac{\epsilon_3{\cdot}k_2 }{ \sigma_{2}}\right)
 	\end{align}
Each term in the expression above equals to the action of a first-order differential operator, which can be written in terms of a common reduction matrix $M_{\sigma_2-\sigma_0}^{(4)}$ and the canonical coefficients of two prepared forms. That is,
 	\begin{align}
 	N_4=&\frac{-s_{12} s_{1,23}\epsilon_1{\cdot} F_{2}{\cdot} \epsilon_4}{s_{13}} \times \nonumber\\
 	& M^{(4)}_{\sigma_2-\sigma_0}\Big(\epsilon_3{\cdot}k_1 a^{(\sigma_0)}_{0,0}-\epsilon_3{\cdot}k_1 a^{(\sigma_2)}_{0,0}-\epsilon_3{\cdot}k_2 a^{(\sigma_2)}_{0,0} \Big)\left.\left[\mathcal D_1^{(1)} {1\over \sigma_0-1}\right] \right|_{\sigma\rightarrow 0},
 \end{align}
 where $\mathcal D_1^{(1)}=\partial^{0,1} -\frac{ s_{12}}{s_{13}}\partial^{1,0}$, $M_{\sigma_2-\sigma_0}^{(4)}$ has been computed previously in (\ref{eq:Mexample4}) and $a^{(\sigma_r)}_{0,0}$ is given by (\ref{eq:valA1}). Explicitly, they read
 \begin{align}
 	M^{(4)}_{\sigma_2-\sigma_0}=-{s_{13}\over s_{1,23} },&&a^{(\sigma_2)}_{0,0}=-{1\over s_{12}}, &&a^{(\sigma_0)}_{0,0}={1\over s_{13}}.
 \end{align}
% Then we get 
% \begin{align}
% 	N_4=-{1\over s_{13}} k_1{\cdot} F_{3}{\cdot} k_2\, \epsilon_1{\cdot} F_{2}{\cdot} \epsilon_4. 
% \end{align}
Evaluating other terms in (\ref{eq:redPfYM4}) similarly, we obtain the 4-point BCJ numerator in the minimal basis
\begin{align}
\label{eq:Amp4BCJ}
  &N(1234)= \epsilon_1{\cdot} F_{2}{\cdot} F_{3}{\cdot} \epsilon_4+\frac{s_{12}}{s_{13}}\epsilon_1{\cdot} F_{3}{\cdot} F_{2}{\cdot} \epsilon_4-\frac{s_{12}}{s_{1,23}}\text{tr}(F_2{\cdot}F_3)\epsilon_1{\cdot}\epsilon_4 \\
  &-\frac{1}{s_{13}}k_1{\cdot} F_{3}{\cdot} k_2\epsilon_1{\cdot} F_{2}{\cdot}\epsilon_4-{1\over s_{13}} k_1{\cdot} F_{2}{\cdot} k_3\, \epsilon_1{\cdot} F_{3}{\cdot} \epsilon_4-\frac{1}{s_{13}s_{1,23}}k_1{\cdot} F_{3}{\cdot} k_2 k_1{\cdot} F_{2}{\cdot} k_3 \epsilon_1{\cdot}\epsilon_4  \,.\nonumber
 \end{align}
As $F_i$ vanishes under the gauge transformation $\epsilon_i\rightarrow k_i$, $N(1234)$ is manifestly gauge invariant in leg $2$ and leg $3$.
At five points, the minimal basis consists of two independent BCJ numerators $N(12345)$ and $N(13245)$. As pointed out before, the numerators computed this way is crossing symmetric and therefore it suffices to compute $N(12345)$ only. This numerator reads
\begin{eqnarray}
\label{eq:AmpYM5}
N(12345) = \oint_{h_1=h_2=0} \sigma_{145}^2\,\overline{\text{PT}}(12345)\text{Pf}\,'[\Psi_{15}(\sigma)] d\Omega_5\,.
\end{eqnarray}
where the dual basis projector $\overline{\text{PT}}(12345)$ can be easily computed from (\ref{eq:ProFac}) and takes the form below
\begin{align}
	\overline{\text{PT}}(12345)=& \left({s_{12}s_{34}s_{1,24}  s_{12,3}\over s_{14} s_{124}}+{s^2_{12}s_{34}s_{13}\over s_{14} s_{134}}\right)\text{PT}(12345)\nonumber\\
	&+\left(\frac{s_{12} s_{13}s_{24} s_{34}}{s_{14} s_{134}}+\frac{s_{12} s_{13}  s_{24} s_{34}}{s_{14} s_{124}}\right)\text{PT}(13245).
\end{align}
The reduced Pfaffian at five points reads \cite{Lam:2016tlk}
\begin{align}
\label{eq:Pfaffian5}
&\text{Pf}\,'[\Psi_{15}(\sigma)] =  \nonumber\\
&\sum_{\gamma\in S_3} \left(-{\epsilon_1{\cdot} F_{\gamma(2)}{\cdot}F_{\gamma(3)}{\cdot}F_{\gamma(4)}{\cdot} \epsilon_5\over \sigma_{1\gamma(2)\gamma(3)\gamma(4)5}}+{\epsilon_1{\cdot} F_{\gamma(2)}{\cdot} \epsilon_5\text{tr}(F_{\gamma(3)}F_{\gamma(4)})\over 4\sigma_{1\gamma(2)5}\sigma_{\gamma(3)\gamma(4)}}\right)+{\epsilon_1{\cdot} \epsilon_5\text{tr}(F_2F_3F_4)\over 2 \sigma_{15}\sigma_{234}}\nonumber \\
&\nonumber\\
& +{\epsilon_1{\cdot} \epsilon_5\text{tr}(F_3F_2F_4)\over 2\sigma_{15}\sigma_{324}}+{ \sum_{\gamma\in S_3} \left({\epsilon_1{\cdot} F_{\gamma(2)}{\cdot}F_{\gamma(3)}{\cdot} \epsilon_5C_{(\gamma(4))}\over \sigma_{1\gamma(2)\gamma(3)5}}-{\epsilon_1{\cdot} F_{\gamma(2)}{\cdot} \epsilon_5C_{(\gamma(3))}C_{(\gamma(4))}\over 2\sigma_{1\gamma(2)5}}\right)}\nonumber\\
&-\epsilon_1{\cdot} \epsilon_5\left({\text{tr}(F_2F_3)C_{(4)}\over \sigma_{15}\sigma_{23}}+{C_{(2)}\text{tr}(F_3F_4)\over \sigma_{15}\sigma_{34}}+{\text{tr}(F_2F_4)C_{(3)}\over \sigma_{15}\sigma_{24}}-{C_{(2)}C_{(3)}C_{(4)}\over \sigma_{15}}\right)  \,.
\end{align}
Like the four-point case, we also consider the characteristic term below as an example here and the rest can all be computed in the same way,
\begin{align}\label{eq:N5example}
N_5 &= \oint_{h_1=h_2=0} \sigma_{145}^2~\overline{\text{PT}}(12345){  \epsilon_1{\cdot}F_2{\cdot}F_3{\cdot}\epsilon_5 C_{(4)} \over \sigma_{1235}} d\Omega_5\,.
\end{align}
Taking $\sigma_1\rightarrow0,\sigma_4\rightarrow1,\sigma_5\rightarrow\infty$, we are only left with two variables $\sigma_2$ and $\sigma_3$. The scattering equations and the denominators above are homogenized with an auxiliary variable $\sigma_0$. This process leads to
\begin{align}
C_{(4)}={s_{14}\over -\sigma_0}+{s_{24}\over \sigma_2-\sigma_0}+{s_{34}\over \sigma_3-\sigma_0}\,,~~~~~~ {1\over \sigma_{1235}} = {1\over \sigma_2 (\sigma_2-\sigma_3)}\,.
\end{align}
There are no other denominators in (\ref{eq:N5example}) after gauge fixing. 
The factor $C_{(4)}$ can now be replaced by a matrix denoted as $M_{C_{(4)}}^{(5)}$, which is a sum over three reduction matrices, each corresponding to the denominator of one of its three terms. These reduction matrices can all be constructed following the discussions in the previous section. This matrix $M_{C_{(4)}}^{(5)}$ reads
\begin{align}
\label{eq:c453}
M_{C_{(4)}}^{(5)}&\equiv\left(
\begin{array}{cc}
{-s_{1,34} s_{124} k_2{\cdot}F_4{\cdot}k_3+s_{234} s_{34}k_1{\cdot}F_4{\cdot}k_2\over s_{234}s_{24}s_{34}}& -\frac{ s_{13} s_{134}k_2{\cdot}F_4{\cdot}k_3}{s_{234}s_{24}s_{34}} \\
-\frac{ s_{12} s_{124}k_3{\cdot}F_4{\cdot}k_2}{s_{234}s_{24}s_{34}}  & {-s_{1,24} s_{134} k_3{\cdot}F_4{\cdot}k_2+s_{234} s_{24} k_1{\cdot}F_4{\cdot}k_3\over s_{234}s_{24}s_{34}}\\
\end{array}
\right)\,
\end{align}
Another common reduction matrix $M_{\sigma _2-\sigma _3}^{(5)}$ renders  (\ref{eq:N5example}) a prepared form. This reduction matrix reads
\begin{align}\label{eq:M235}
M_{\sigma _2-\sigma _3}^{(5)}&=\left(
\begin{array}{cc}
 \frac{\left(s_{12} s_{14}+s_{1,34} s_{23}\right) s_{124}}{s_{1,234}s_{23} s_{123}} & \frac{s_{13} \left(s_{14}-s_{23}\right)s_{134,2}}{s_{1,234} s_{23} s_{123}} \\
 \frac{s_{12} \left(s_{14}-s_{23}\right) s_{124}}{s_{1,234} s_{23} s_{123}} & \frac{\left(s_{13} s_{14}+s_{1,24}s_{23}\right) s_{134,2}}{s_{1,234} s_{23} s_{123}} \\
\end{array}
\right).
\end{align}
The canonical $a$-coefficients for the prepared form with $h_0=\sigma_2$ are
\begin{align}
\label{eq:asigma2}
a_{010}^{(\sigma_2)}=\frac{1}{s_{12}s_{124}}, &&a_{100}^{(\sigma_2)}=0.
\end{align}
Hence (\ref{eq:N5example}) is given by the action of the differential operator as follows
\begin{align}
N_5 &=-{\epsilon_1{\cdot}F_2{\cdot}F_3{\cdot}\epsilon_5\over 2}\times\\
&\sum_{i,j=1}^2\Big( M^{(5)}_{C_{(4)}}M^{(5)}_{\sigma_2-\sigma_3}\Big)_{i,j}a_{\gamma_i(0),\gamma_i(1),0}^{(\sigma_2)}\left.\left[\mathcal D^{(3)}_j {\sigma_{145}~\overline{\text{PT}}(12345)J_5(\sigma) \over (\sigma_0-1)} \right] \right|_{\sigma \rightarrow 0} \,,\nonumber
\end{align}
where $\mathcal D_j^{(3)}$ is given in \eqref{eq:DiffOPFiveHighest} and $\gamma_i\in S_2$. 
Plugging in the expressions \eqref{eq:DiffOPFiveHighest}, \eqref{eq:c453}, \eqref{eq:M235} and \eqref{eq:asigma2}, we obtain
\begin{align}\label{eq:Five2}
 N_5&=\frac{k_1{\cdot}F_4{\cdot}k_3}{s_{14}}\epsilon_1{\cdot}F_2{\cdot}F_3{\cdot}\epsilon_5-\frac{s_{34}k_1{\cdot}F_4{\cdot}k_2}{s_{14} s_{124}}\epsilon_1{\cdot}F_2{\cdot}F_3{\cdot}\epsilon_5. 
\end{align}
%It is easy to see that the gauge invariances for leg 2, 3, 4 are all obvious. This term contributes to the BCJ numerator $N(12345)$. Similarly for the other terms in the Pfaffian expansion, we use the dual basis and reduction matrix to evaluate the BCJ numerator directly and get 
All other terms in (\ref{eq:Pfaffian5}) are treated similarly and the explicit expression for the five-point BCJ numerator reads
\begin{align}
	 &N(12345) =  \nonumber\\
	 &=\left(k_1{\cdot}F_4{\cdot}k_3\,-\frac{s_{34} }{s_{124}}k_1{\cdot}F_4{\cdot}k_2\right)\,{\epsilon_1{\cdot}F_2{\cdot}F_3{\cdot}\epsilon_5\over s_{14}}+\frac{ 1}{ s_{14}}k_1{\cdot}F_2{\cdot}F_3{\cdot}F_4{\cdot}k_1 \epsilon_1{\cdot}\epsilon_5\nonumber\\
	 &+\left(\frac{1}{ s_{134}}k_1{\cdot}F_2{\cdot}k_{34}k_1{\cdot}F_3{\cdot}F_4{\cdot}k_1+\frac{ 1}{s_{124}}k_{12}{\cdot}F_3{\cdot}k_{4}k_1{\cdot}F_2{\cdot}F_4{\cdot}k_1\right){\epsilon_1{\cdot}\epsilon_5\over s_{14}}\nonumber\\
	 %%%%%%%%%%%%%%%%%%%%%%%
	 &+\left({s_{12} s_{23}+s_{14} s_{2,34}\over s_{124} s_{14} s_{134}}k_1{\cdot}F_3{\cdot}F_4{\cdot}k_1 +\frac{s_{12} s_{23}+s_{12,3} s_{2,34}}{s_{25}s_{34}}\text{tr}(F_3{\cdot}F_4) \right)\epsilon_1{\cdot}F_2{\cdot}\epsilon_5\nonumber\\
	 &-\left(k_1{\cdot}F_3{\cdot}F_4{\cdot}k_2+{s_{1,24}\over s_{14} }k_2{\cdot}F_3{\cdot}F_4{\cdot}k_1+k_2{\cdot}F_3{\cdot}F_4{\cdot}k_{2}\right){\epsilon_1{\cdot}F_2{\cdot}\epsilon_5 \over s_{124}}\nonumber\\
	 %%%%%%%%%%%%%%%%%%%%%%%%%%%%%%%
	 &+\left({\left(s_{14}-s_{23}\right) s_{34}\over s_{124} s_{14} }k_1{\cdot}F_2{\cdot}F_4{\cdot}k_1-k_1{\cdot}F_2{\cdot}F_4{\cdot}k_3-{1\over s_{14} }k_1{\cdot}F_2{\cdot}k_3\,k_1{\cdot}F_4{\cdot}k_3\right) {\epsilon_1{\cdot}F_3{\cdot}\epsilon_5\over s_{134}}\nonumber\\
	 %%%%%%%%%%%%%%%%%%%%%%%%%
	 &+\left(k_1{\cdot}F_2{\cdot}F_3{\cdot}k_4+\frac{s_{12}}{s_{14}}k_{14}{\cdot}F_2{\cdot}F_3{\cdot}k_4-\frac{ s_{1,24}}{ s_{14} }k_1{\cdot}F_2{\cdot}k_3\,k_1{\cdot}F_3{\cdot}k_4\right){\epsilon_1{\cdot}F_4{\cdot}\epsilon_5\over s_{124}}\nonumber\\
	 &+\left(\frac{ s_{12}}{s_{14} s_{134}}k_3{\cdot}F_2{\cdot}k_4\,k_1{\cdot}F_3{\cdot}k_4-\frac{1}{s_{134}}k_1{\cdot}F_2{\cdot}k_4\,k_1{\cdot}F_3{\cdot}k_4\right){\epsilon_1{\cdot}F_4{\cdot}\epsilon_5\over s_{124} }\nonumber\\
	 %%%%%%%%%%%%%%%%%%%%%
	 &+{s_{12} \over  s_{14} s_{134}}k_1{\cdot}F_4{\cdot}k_3\,\epsilon_1{\cdot}F_3{\cdot}F_2{\cdot}\epsilon_5+{1\over s_{124}}k_{12}{\cdot}F_3{\cdot}k_4\,\epsilon_1{\cdot}F_2{\cdot}F_4{\cdot}\epsilon_5\nonumber\\
	 &+\left({ \left(s_{14}-s_{23}\right) \over  s_{134}}k_1{\cdot}F_3{\cdot}k_4+k_2{\cdot}F_3{\cdot}k_4\right){s_{12}\epsilon_1{\cdot}F_4{\cdot}F_2{\cdot}\epsilon_5\over s_{14} s_{124}} \nonumber\\
	 &-\left({s_{1,24}s_{13,4} \over  s_{14} }k_1{\cdot}F_2{\cdot}k_{3}-s_{13}k_1{\cdot}F_2{\cdot}k_{4}-{s_{12}s_{13,4}\over s_{14} }k_3{\cdot}F_2{\cdot}k_{4}\right){\epsilon_1{\cdot}F_4{\cdot}F_3{\cdot}\epsilon_5\over s_{124}s_{134}}\nonumber\\
	 %%%%%%%%%%%%%%%%%
	 &+{1\over s_{134}}k_1{\cdot}F_2{\cdot}k_{34}\,\epsilon_1{\cdot}F_3{\cdot}F_4{\cdot}\epsilon_5-\epsilon_1{\cdot}F_2{\cdot}F_3{\cdot}F_4{\cdot}\epsilon_5%%
-\frac{s_{12,3}}{s_{124}}\epsilon_1{\cdot}F_2{\cdot}F_4{\cdot}F_3{\cdot}\epsilon_5%%
\nonumber\\
&
-\frac{s_{12}}{s_{134}}\epsilon_1{\cdot}F_3{\cdot}F_4{\cdot}F_2{\cdot}\epsilon_5+\frac{s_{12} s_{34}}{s_{124} s_{14}}\epsilon_1{\cdot}F_4{\cdot}F_2{\cdot}F_3{\cdot}\epsilon_5+\frac{s_{12} s_{13,4}}{s_{134} s_{14}}\epsilon_1{\cdot}F_4{\cdot}F_3{\cdot}F_2{\cdot}\epsilon_5. %%
\end{align}
The other numerator is obtained by the permuting indices $$N(13245)=N(12345)|_{2\leftrightarrow 3}.$$
These non-local BCJ numerators are compact and again manifestly gauge invariant in the leg 2,3 and 4, due to the existence of the $F_i$ factors.

%As always, the contribution of this term $I_2$ to the BCJ numerator in the minimal basis is computed by the following inner products respectively
%\begin{align}
%\langle \overline{\text{PT}}_{12345}, I_2 \rangle_{CHY}  \,, ~~~~~~ \langle \overline{\text{PT}}_{13245}, I_2 \rangle_{CHY} \,,
%\end{align} 
%which in practice is to pick up the coefficients of the Parke-Taylor factors in \newref{eq:Five2}.

%All other terms in \newref{eq:Pfaffian5} can be dealt with in the same fashion. Since the two terms we have discussed in detail are representative enough, the remaining ones possess no conceptual novelty and we present the final expression of them in Appendix \ref{sec:5pointFinal}. 

\section{Conclusion and Outlooks}
In this paper, we have developed a systematic method to construct the BCJ numerators, starting from the CHY forms of scattering amplitudes and using the differential operator proposed in our previous work \cite{Chen:2016fgi}.
This method is based on the key observation that the number of canonical coefficients in such a differential operator is always $(n-3)!$ for a $n$-point CHY form, independent of the order of the operator.
In the process of solving for the canonical coefficients, we have built the reduction matrices to simplify the differential operator and improve the efficiency of computation. The reduction matrices are also universal for all the theories. In the end, as we have demonstrated, we always arrive at the prepared forms for which the coefficients are solved analytically in \cite{Chen:2017edo}. 

Both the BCJ numerators and the amplitudes obtained this way enjoy the manifest gauge invariance in $(n-2)$ out of the $n$ external legs, and their final expressions are always of factorized forms.
%We can keep the gauge invariance manifest  for both the amplitude and the BCJ numerator. Furthermore, the evaluated result is of the same factorized form as the CHY integrand.  
%
%Our method is based on the differential operator form of the multi-residue integral which is introduced in our  former work in \cite{Chen:2016fgi}. The coefficients of the differential operator is determined by the scattering equations and denominator $h_0$ of the CHY integrand.  A key observation is  that, after solving all the local duality equations from  the polynomial in the scattering equations, there are only  $(n-3)!$ canonical a-coefficients.  Moreover, the number is independent of the order of $h_0$. Then we find, in evaluating these canonical a-coefficients, each factor $\sigma_{ij}$ in $h_0$ can be replaced by the reduction matrix which is also  irrelevant with the theory.   The reduction is end with a prepare form which is theory dependent but of quite simple form for general n-point amplitude\cite{Chen:2017edo}. Combining with the theory independent reductive matrix and the theory dependent prepare form, we can evaluate the a-coefficients efficiently. The values of these $(n-3)!$ canonical a-coefficients are just the BCJ numerators in the unit basis.   All the other a-coefficients can be deduced by the local duality conditions from the scattering equations. After getting the differential operator, the CHY evaluating becomes an easy task. 
It is hopeful to formulate a closed form for the reduction matrix in future works, since a given reduction matrix $M_{\sigma_{i}}$ is determined only by the factor $\sigma_{i}$ and the polynomial scattering equations.  Moreover, the polynomial scattering equations have nice combinatoric structures to exploit, which may allow us to produce analytical results for a general reduction matrix. As discussed in \cite{Bosma:2016ttj}, the polynomial scattering equations is a Macaulay H-basis. This property may be helpful to prove the observation of the number $(n-3)!$ of independent $a$-coefficients.

In this paper, the concept of dual basis is adopted to extract the non-local BCJ numerators in the minimal basis, in which the non-local propagators can be removed using the BCJ relations. It is conceivable that similar projectors can be constructed for the local BCJ numerators as well. It can be expected that such projectors are constructed recursively, which may point to novel algebraic structures \cite{Dri89,dascalescu2000hopf,Fu:2016plh,Duhr:2012fh}.

%Besides the algebraic properties, it is also important to investigate the geometric aspects of the BCJ numerators. The geometry of the scalar amplitudes, namely the cubic structures, is scrutinized in the language of associahedron in \cite{Arkani-Hamed:2017mur,Mizera:2017cqs}. There can be ways to dress these geometries with numerators encoding the kinematic information, so that the dressed geometries capture the properties of other theories, such as gauge theory and gravity \cite{Arkani-Hamed:2013jha,Arkani-Hamed:2017tmz,Arkani-Hamed:2017fdk}.

Our method can be easily generalized to loop levels. The one-loop scattering equations and the prepared forms have been studied in \cite{Chen:2017edo}. We expect the canonical coefficients can also be found at one loop and then the construction of reduction matrix is expected to be straightforward.  
%Even if the generalized CHY forms for higher-loop amplitudes remain unknown, the dimension of the basis for loop-level BCJ numerators can still be calculated from the scattering equations.
%Moreover, even if we do not know the higher loop generalized CHY form, it is still possible to count the number of the BCJ basis only from the scattering equations.
Another future direction is to carry our method over to string theory. String amplitudes, written in the forms of worldsheet integrals, have a number of features in common with the CHY-like integrals. It is reasonable to hope that there exist similar differential operators and even reduction matrices which can help us evaluate those worldsheet integrals efficiently while preserving the factorized form. 
% Another direction is extend our method to string theory amplitude. Is there also a reductive matrix for each factor in the string world sheet integral? We will leave all of these problem in future work.

\begin{acknowledgements}
G.C. and T.W. thank H. Johansson for useful discussion and kind suggestions. The research of G.C. is supported in part by the Knut and Alice Wallenberg Foundation under grant KAW 2013.0235, and the Ragnar S\"{o}derberg Foundation under grant S1/16.
G.C.  has been supported in parts by  NSF of China Grant under Contract 11405084, the Open Project Program of State Key Laboratory of Theoretical Physics, Institute of Theoretical Physics, Chinese Academy of Sciences (No.Y5KF171CJ1). T.W. thanks the China Scholarship Council for support (File No. 201706190098).
\end{acknowledgements}

\appendix
\bibliographystyle{JHEP}      % basic style, author-year citations

%\bibliographystyle{spphys}
\bibliography{ScatEq}

\end{document}